\documentclass[a4paper,11pt]{article}
\usepackage{float}
\usepackage[pdftex]{graphicx}
\usepackage{subcaption}
\usepackage[T1]{fontenc}
\usepackage{lmodern}
\usepackage{slashed}
\usepackage[utf8]{inputenc}
\usepackage[english]{babel}
\usepackage{microtype}
\usepackage{cite}
\usepackage{amsmath,amssymb,amsfonts,amsthm}
\usepackage{mathtools,mathrsfs,calligra,aurical}
\usepackage[nottoc,notlot,notlof]{tocbibind}
\usepackage{upgreek}
\usepackage{mathtools}
\numberwithin{equation}{section}
\allowdisplaybreaks
\usepackage[all]{xy}
\usepackage{xcolor}
\usepackage{graphicx}
\graphicspath{{images/}}
\usepackage{geometry}
\usepackage[toc,page]{appendix}
\usepackage{hyperref}
\usepackage[normalem]{ulem}

\usepackage{bm}
\usepackage{ragged2e}
\usepackage{appendix}
\usepackage{slashed}
\usepackage{bbold}
\usepackage{cancel}

\definecolor{blue-violet}{rgb}{0.54, 0.17, 0.89}
\definecolor{PineGreen}{cmyk}{0.92, 0, 0.59, 0.25}
\definecolor{YellowOrange}{cmyk}{0, 0.42, 1, 0}
\definecolor{orange}{rgb}{0.95, 0.5, 0.1}

\DeclareMathAlphabet{\mathpzc}{OT1}{pzc}{m}{it}

\begin{document}

\begin{titlepage}
\begin{flushright}
\par\end{flushright}
\vskip 0.5cm
\begin{center}
\textbf{\LARGE \bf  Plasma-Plasma Third Order Phase Transition from type IIB Supergravity}\\
\vskip 5mm

\vskip 1cm

\large {\bf Andr\'{e}s Anabal\'{o}n}\footnote{anabalo@gmail.com} and \large {\bf Julio Oliva}\footnote{ julioolivazapata@gmail.com}

\vskip .5cm 

{\textit{Departamento de F\'isica, Universidad de Concepci\'on, Casilla, 160-C, Concepci\'on, Chile.}}
\end{center}
\begin{abstract}
{We show that the planar, charged black hole in AdS, dual to the strongly coupled Quark-Gluon Plasma thermal state of large $N$, $SU(N)$, $\mathcal{N}=4$ super Yang-Mills at finite chemical potential undergoes a third-order phase transition in the grand canonical ensemble to a hairy black hole of type IIB supergravity. The hairy phase is another strongly coupled fluid with a conformal equation of state and can be interpreted as another kind of Quark-Gluon plasma. This new Quark-Gluon plasma has less entropy and, therefore, seems to characterize some form of smooth hadronization.  
The locus of the transition in terms of the ``Baryon'' chemical potential, $\mu$, and the temperature, $T$, is $\mu=2\pi T$.}
\end{abstract}

\vfill{}
\vspace{1.5cm}
\end{titlepage}

\setcounter{footnote}{0}

\section{Introduction and Discussion}
In an experimental breakthrough, a new phase of matter, the strongly coupled Quark-Gluon Plasma (\textbf{QGP}) was created in ultrarelativistic heavy ion collisions \cite{STAR:2005gfr, ALICE:2010suc}, in which the quarks and gluons are not confined but also not free. State of the art measurements shows that its speed of sound squared is $c^2_s\approx 0.241$ (in units of the speed light squared) at effective temperatures of $T_{eff}\approx 219$ [MeV] ($T_{eff}$ is $1/3$ of the mean transverse momentum of the collisions)  \cite{CMS:2024sgx}. This is in very good agreement with lattice Quantum Chromo Dynamics (\textbf{QCD})\cite{HotQCD:2014kol}. Furthermore, at zero baryon density, lattice QCD indicates that the confinement/deconfinement phase transition is a
 crossover \cite{Aoki:2006we} \footnote{In the context of lattice QCD, because of the computational resolution, phase transitions of order higher than two are called crossovers. We thank Fabrizio Canfora for this remark.}.  
 
 The many future experiments probing the QGP are in need of a theoretical benchmark to compare with. In particular regarding on the details of the possible phase transitions that can be found at high temperatures and high Baryon chemical potentials, situations in which lattice QCD face the sign or complex action problem \cite{Nagata:2021ugx}.
 
 QCD is asymptotically free, hence the QGP at high enough energies should yield a gas of non-interacting particles. In the regime where the QGP is still strongly coupled with a speed of sound close to the conformal value $c^2_s= \frac{1}{3}$, one can hope to provide a good description of certain aspects of the plasma using a conformal field theory, by means of holography \cite{Maldacena:1997re}. A remarkable example of this program is the bound on the shear-viscosity to entropy density ratio \cite{Kovtun:2004de}. Further insights on the behaviour of the QGP might arise from phase diagram of the phase transitions, where critical exponents and the order of the phase transition characterize a universality class.

Holography allows to describe strongly coupled quantum field theories in the large $N$ limit. In particular, since its inception, it was remarked that the well-known Hawking-Page phase transition for black holes in asymptotically anti-de Sitter spacetimes \cite{Hawking:1982dh} can describe a confinement/deconfinement phase transition at finite temperature on the strongly coupled large $N$, $SU(N)$, $\mathcal{N}=4$ Super Yang-Mills thermal quantum field theory when the geometry of the spacetime is $S^3\times S^1$ \cite{Witten:1998zw}. The black hole is dual to the deconfined state and thermal AdS is dual to the confined phase. When the horizon of the black hole is $\mathbf{R}^3$ the dual thermal quantum field theory is defined on $\mathbf{R}^3\times S^1$ and the expectation value of the energy momentum tensor yields a conformal fluid with relativistic Stefan-Boltzmann behaviour, which motivates its identification with the QGP. However, the Hawking-Page phase transition does not seem to be relevant for field theories on thermal Minkowski $\mathbf{R}^3\times S^1$. This follows from the fact that the Gibbs free energy is everywhere negative in this case.

When the field theory lives in $\mathbf{R}^3\times S^1$, the QGP thermal state is dual to the five-dimensional asymptotically AdS, charged, planar black hole solution of the Einstein-Maxwell theory in five dimensions with a negative cosmological constant. This purely electrically charged black hole is also a solution of the STU model of the maximally supersymmetric type IIB gauged supergravity with its three $U(1)$ gauge fields equal to each other\footnote{Each of the three gauge fields induce an expectation value for an R-symmetry current in the field theory with baryonic number densities $N_1, N_2, N_3$.}. Hence, the grand canonical ensemble is characterized by the black hole densities of: energy ($\rho$), entropy ($S$), baryonic number  $N_1=N_2=N_3=B$ and Gibbs free energy ($G$). The intensive parameters are the black hole temperature $T$ and chemical potential conjugate to $B$, $\mu$, which can be combined in the dimensionless quantity $\psi=\frac{\mu}{2 \pi T}$. A straightforward calculation yields\footnote{
In natural units, $\hbar=c=k_B=1.$} 
\begin{align}\label{phase1}
\rho&=\sigma T^4 \left(1+12\psi^2+24\psi^4+(1+8\psi^2)^{3/2}\right)\, ,\\ B&=\sigma \frac{4 T^3}{3\pi}\left(1+(1+8\psi^2)^{1/2}+4\psi^2\right)\psi\, ,\\
S&=\sigma \frac{4 T^3}{3}\left(1+6 \psi^2+(1+2 \psi^2)(1+8\psi^2)^{1/2}\right)\, ,\\ G(T,\mu)&=\rho-T S-\sum^3_{i=1} N_i\mu=-\frac{\rho}{3}\, ,
\end{align}
with $\sigma=\frac{3\pi^4 L^3}{4\kappa}$, $L$ is the AdS radius and $\kappa=8\pi G_N$ is the reduced Newton constant in five dimensions, field theory variables are obtained with the identification $\frac{L^3}{\kappa}=\frac{N^2}{4\pi^2}$. The AdS/CFT dictionary allows to identify $\rho$ and $P=\frac{\rho}{3}$ as the expectation values of the dual QGP energy momentum tensor. This is the standard dual to the QGP plasma at finite chemical potential. It yields a strongly coupled fluid with speed of sound $c^2_s=\frac{\partial P}{\partial \rho}=\frac{1}{3}$. This phase is globally stable stable as the Gibbs free energy is everywhere concave

\begin{align}\label{stability}
\frac{\partial^2 G}{\partial \mu^2}\leq 0 \qquad \frac{\partial^2 G}{\partial T^2}\leq 0\qquad \frac{\partial^2 G}{\partial \mu^2} \frac{\partial^2 G}{\partial T^2}-\left(\frac{\partial^2 G}{\partial T \partial \mu} \right)^2\geq 0 \, ,
\end{align}
In \eqref{stability} the inequalities are all saturated only at $(T,\mu)=0$. Therefore, any possible instability of this phase is quite unexpected.

In this paper we report the existence of a third order phase transition in the infinite volume limit. The transition is to another black hole, hence describes another QGP of $SU(N)$, $\mathcal{N}=4$ Super Yang-Mills, with equation of state $P=\frac{\rho}{3}$ and therefore conformal speed of sound in the $AdS_5\times S^5$ compactification of IIB supergravity. This other black hole is a limit of the solutions obtained in \cite{Behrndt:1998jd}. The black hole has a running dilaton $\phi$, with mass saturating the Breithenloner-Freedman bound in five dimensions. It yields a non-trivial expectation value to a bi-linear of the scalars in the gauge theory however, it is also asymptotically AdS with the same boundary conditions, characterized by $(T,\mu)$, than the standard solution above. It has thermodynamical variables 

\begin{align}\label{phase2}
\rho_{\phi}&=32 \sigma T^4(1+\psi^2)\psi^2\, , \\ N_{1\phi}&=\sigma \frac{32 T^3\psi^3 }{3\pi}=N_{2\phi}\, ,\\
N_{3\phi}&=\sigma \frac{32 T^3\psi }{3\pi}\, ,\\
S_{\phi}&=\sigma \frac{64}{3}\psi^2 T^3\, ,\\ 
G_{\phi}(\mu,T)&=\rho_{\phi}-T S_{\phi}-\sum^3_{i=1} N_{i\phi}\mu=-\frac{\rho_{\phi}}{3}\, .
\end{align}
This black hole does not allow for a $T=0$ limit when $\mu\neq 0$, at fixed boundary conditions. At $\mu=0$ it yields some of supersymmetric distributions of D3-branes, with no horizon, of \cite{Freedman:1999gk}. From \eqref{stability}, it follows that this phase is locally stable provided that $\psi> 2^{-1/2}$.

There is a phase transition at $\psi=1$. At this point we find that
\begin{equation}
G-G_{\phi}=\frac{2^8\sigma}{3^4} (\psi - 1)^3+O(\psi - 1)^4\, .
\end{equation}
Hence, we have a line of third order phase transitions separating two possible QGP, both with speed of sound, $c_s^2=\frac{1}{3}$. Our findings are collected in the ($T$,$\mu$) phase diagram of figure 1.

\begin{figure}[H]
\centering
\includegraphics[scale=0.8]{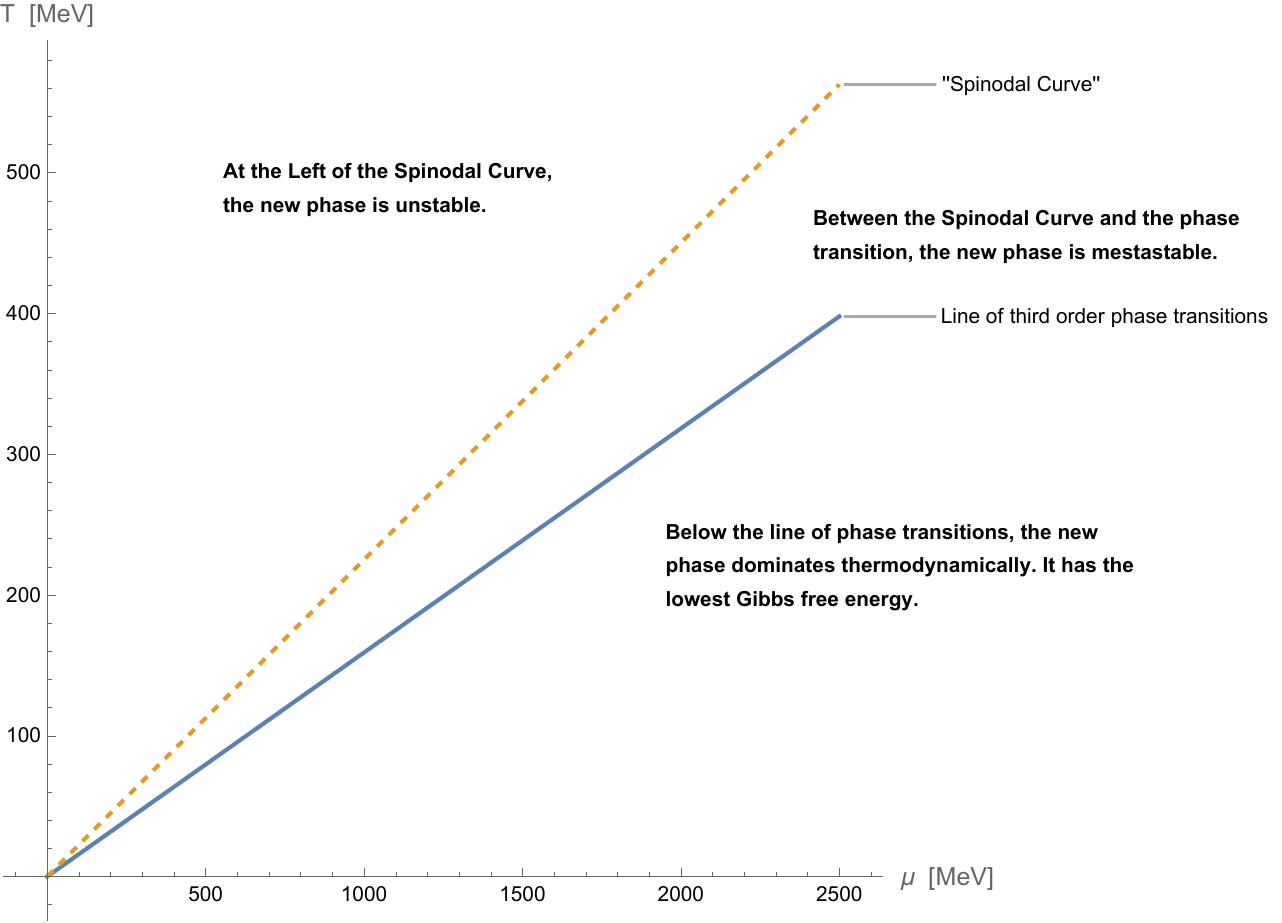}
\caption{The phase diagram in the $(T, \mu)$ plane for the QGP from type IIB supergravity. There is a line of third order phase transitions at $T=\frac{\mu}{2\pi} $. The new QGP dominates below that line. The normal QGP dominates above that line. In the region above the dashed line, the new phase is unstable. The determinant of the Hessian of the Gibbs free energy of the new QGP vanishes at the dashed line. The new phase stop existing at $T=0$ for $\mu\neq 0$. At $\mu=0$ is a stack of D3 branes that has no horizon.}
\label{fig1}
\end{figure}
The phase diagram diagram of figure 1 is exact for the black holes of type IIB supergravity and for its strongly coupled QFT dual in the large N limit. It yields a qualitative picture of the critical behaviour of QCD in the regions where $c_s^2\approx 0.3$, namely temperatures $T\approx 370$ [MeV] at zero chemical potential \cite{HotQCD:2014kol}. Furthermore, using lattice QCD, its critical point has been excluded from the region $\mu\leq \pi\, T$, $\mu < 300 [MeV]$ \cite{Vovchenko:2017gkg, Borsanyi:2020fev, Borsanyi:2021sxv}. Our results suggest that QCD should have a critical point around $(T_c,\mu_c) \approx (T_c,2\pi T_c)$, with a line of third order phase transitions emerging from that point with positive slope. 

An analogue phenomena happens for spherical and hyperbolic black holes and will be presented elsewhere \cite{AO2024}.

The details of the calculation are given below. We present the supergravity truncation with our conventions and then the black solution that produce the new plasma phase.

\section{The supergravity truncation in five dimensions}

The two plasma phases are solutions of the STU model. In our conventions the action principle is 
\begin{equation}
I_0=\frac{1}{2\kappa}\int\sqrt{-g}\left(  R-\frac{\left(
\partial\Phi_{1}\right)  ^{2}}{2}-\frac{\left(\partial\Phi_{2}\right)  ^{2}}{2}+\sum_{i=1}^{3}4L^{-2}X_i^{-1}-\frac{1}{4}X_{i}^{-2}(F^{i})^{2}+\frac{1}{4}\epsilon^{\mu \nu \rho \sigma \lambda}A^1_{\mu}F^2_{\nu \rho}F^3_{\sigma \lambda} \right)  d^{5}x,
\label{LSTU}%
\end{equation}
where $F_{i}=d\bar{A}_{i}$, $X_{i}=e^{-\frac{1}{2}\vec{a}_{i}\cdot
\vec{\Phi}}$, $\vec{\Phi}=\left(  \Phi_{1},\Phi_{2}\right)  $ and%
\begin{equation}
\vec{a}_{1}=\left(  \frac{2}{\sqrt{6}},\sqrt{2}\right),\qquad\vec{a}_{2}=\left(  \frac{2}{\sqrt{6}},-\sqrt{2}\right),\qquad\vec{a}_{3}=\left(  -\frac{4}{\sqrt{6}},0\right).
\end{equation}
We are interested in purely electric solutions, where it is
consistent to truncate the axions to zero. The Lagrangian \eqref{LSTU} can be
obtained from the compactification of ten dimensional type IIB supergravity over the $S^5$ \cite{Cvetic:1999xp}.

It is well-known that the Einstein-Maxwell-AdS system with a Chern-Simmons term is recovered when the scalar fields are all zero and the gauge fields are all equal. The electrically charged black hole solution with conformal boundary metric 
\begin{equation} \label{bc1}
ds_{\gamma}^2=\gamma_{a b}dx^adx^b=-dt^2+dy^2+dz^2+dw^2\, ,    
\end{equation}
at temperature $T$ and boundary value for the gauge fields
\begin{equation}\label{bc2}
A^1_t=A^2_t=A^3_t=\mu L \, ,
\end{equation}
is the usual planar Einstein-Maxwell ``Reissner–Nordstr{\"o}m''-AdS black hole, see for instance \cite{Chamblin:1999tk}. Then, a straightforward calculation yields \eqref{phase1}.

We will show below that there is another black hole solution to this system with the same boundary conditions. To this end we study a known solution \cite{Behrndt:1998jd} with  gauge fields that are not everywhere equal and then impose the boundary conditions \eqref{bc1} and \eqref{bc2}. This procedure yields the new plasma phase \eqref{phase2}.

\section{The Hairy Black Hole} 
\begin{figure}[H]
\centering
\includegraphics[scale=0.5]{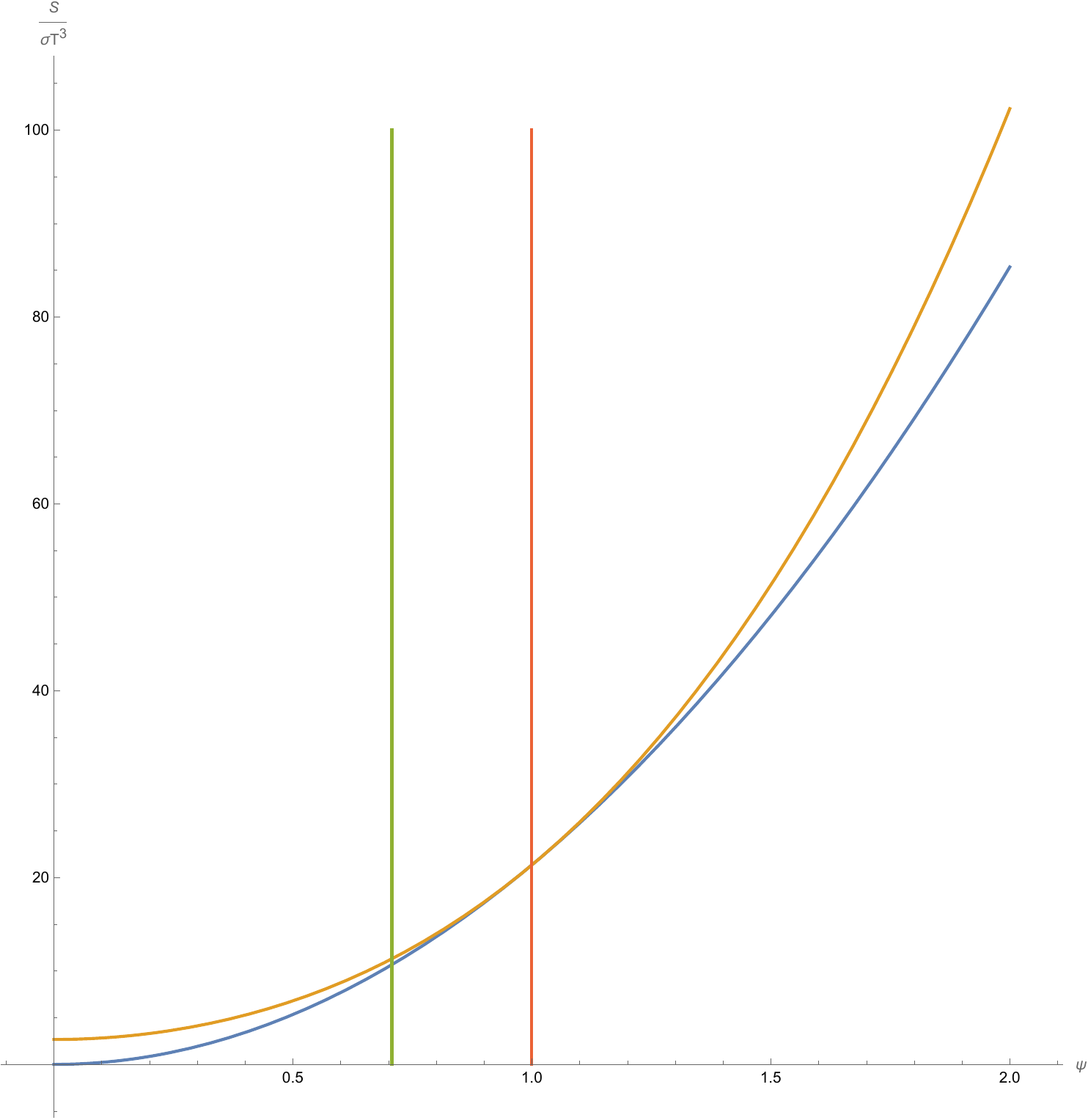}
\caption{The dimensionless entropy density $\frac{S}{\sigma T^3}$ versus $\psi$. The blue line represents the phase with the scalar condensate and the yellow line without it. The phase transition happens in the red line $\psi=1$. The green line indicates the region of the parameters, $\psi=2^{-1/2}$ that onsets the thermodynamical instability of hairy branch.}
\label{fig2}
\end{figure}

These solutions are a particular case of the electrically charged black hole solutions of \cite{Behrndt:1998jd}, which oxidize to spinning D3 branes in 10 dimensions \cite{Cvetic:1999xp}.  The solution is (in a convenient coordinate system)
\begin{equation}
ds^2 =\Omega(x)\left(-F(x) L^2 dt^2+dy^2+dz^2+dw^2+\frac{dx^2}{4 x^2 (x-1) F(x) \eta}\right)\, ,
\end{equation}
\begin{equation}
\Omega(x)=\frac{x^{2/3}\eta}{x-1}\, ,\qquad F(x)=L^{-2}-\frac{(-1+x)^2(q_{1}^2-q_{2}^2 x)}{\eta x^2}\, ,
\end{equation}
\begin{equation}
\Phi_1 =\sqrt{\frac{2}{3}}\ln(x)\, ,\qquad \Phi_2=0\, ,
\end{equation}
\begin{equation}\label{gfh}
A^{1}= q_{1}\left(  x^{-1}-x_{0}^{-1}\right) L dt = A^{2}\, ,\qquad
A^{3}= q_{2}\left(  x-x_{0}\right) L  dt\, ,
\end{equation}
where $F(x_0)=0$. The advantage of this coordinate system is that allows to explore the space of parameters at fixed values of the scalar field. When the integration constant, $\eta>0$ then $1<x$. When $\eta<0$ then $x<1$. These two cases are not diffeomorphic to each other as the scalar field is either everywhere positive or negative depending on which case one considers. 
The conformal boundary is located at $x=1$. This is more easily seen with the change of coordinates
\begin{equation}
x=1+\frac{\eta L^2}{r^2}+\frac{2 \eta^2 L^4}{3 r^4}+\frac{\eta^3 L^6}{3 r^6}+O(r^{-8}),\\
\end{equation}
which implies,
\begin{align}
\Omega(x)  &  =\frac{r^{2}}{L^{2}}+O(r^{-4}),\\
-g_{t t}  &  =\Omega(x)F(x)L^2=\frac{r^{2}}{L^{2}}-\frac{m}{r
^2}+O(r^{-4}),\\
g_{r r}  & =\frac{L^{2}}{r^{2}}-\frac{\frac{2}{9}\eta^2 L^6-m L^4}{r
^6}+O(r^{-8}),\\
m &  =-\eta L^{4} (q_{1}^{2}-q_{2}^{2}).\label{mu}
\end{align}
It follows that the conformal boundary metric is \eqref{bc1}. The boundary condition on the gauge fields  \eqref{gfh} allows to fix $q_1$ and $q_2$
\begin{equation}
q_1=\frac{\mu x_0}{x_0-1} \qquad q_2=-\frac{\mu}{x_0-1}\, ,
\end{equation}
which allows to find the parameter $\eta$ 
\begin{equation}
F(x_0)=0\implies \eta=\frac{\mu^2 L^2 (x_0-1)}{x_0} \, . 
\end{equation}
Finally, the definition of the temperature fixes $x_0$ in terms of the physical variables,
\begin{equation}
x_0=\left(\frac{\mu}{2\pi T}\right)^2\equiv \psi^2 \, .
\end{equation}
Hence, we have constructed a hairy black hole solution with the same boundary conditions than \eqref{phase1}, this phase is characterized by the quantities given in \eqref{phase2}. One can see that the solutions with $\psi>1$ have a positive dilaton and the solutions with $\psi<1$ have a negative dilaton. We use holographic renormalization to compute the expectation value of the  dual energy-momentum tensor, \cite{Balasubramanian:1999re, Bianchi:2001kw, Bianchi:2001de},
\begin{equation}\label{T}
\left<T_{t t}\right>=\frac{3 m}{2 L^3 \kappa}\;,\qquad\left<T_{z z}\right>=\left<T_{y y}\right>=\left<T_{w w}\right>=\frac{m}{2 L^3 \kappa}\, .
\end{equation}
The energy density in \eqref{phase2} is the $\left<T_{t t}\right>$ obtained in \eqref{T}. The entropy density is the usual Bekenstein-Hawking area-law
\begin{equation}
S_{\phi}=\frac{\Omega(x_0)^{3/2}}{4G_N}\, ,
\end{equation}
and $N_{1\phi}, N_{2\phi}, N_{3\phi}$ are the normalized electric charges of the hairy black hole. The entropy of the two phases are contrasted in the figure \ref{fig2}. The hairy phase has always less entropy than the non-hairy phase. Therefore, one can think on it as some form of hadronization of the degrees of freedom of the more entropic phase.  

The R-symmetry is broken from $SO(6)$ to $U(1)^3$, and the dilaton induces an expectation value for the bilinear of the six super Yang-Mills scalars $Z_i$ that is invariant under $SO(4)\times SO(2)$ of the form \cite{Witten:1998qj}
\begin{equation}
\left<\mathcal{O}\right>=\frac{1}{N}\left<Tr(2Z_1^2+2Z_2^2-Z_3^2-Z_4^2-Z_5^2-Z_6^2)\right>\, ,
\end{equation}
which changes sign along the blue line of phase transitions of figure 1. 
\newline
\newline
\textbf{Acknowledgements}
Our work is supported in part by the FONDECYT grants 1210635, 1221504, 1230853 and 1242043. 

\newpage

\hypersetup{linkcolor=blue}
\phantomsection 
\addtocontents{toc}{\protect\addvspace{4.5pt}}
\addcontentsline{toc}{section}{References} 
\bibliographystyle{mybibstyle}
\bibliography{bibliografia.bib}

\end{document}